\documentclass[cameraready]{Interspeech}

\usepackage{booktabs}
\usepackage{multirow}
\usepackage[table]{xcolor}
\usepackage{colortbl}
\usepackage{cleveref}
\usepackage{amsmath}

\title{Evolution Strategy-Based Calibration for Low-Bit Quantization of Speech Models}

\author[orcid=0009-0008-1383-6460]{Lucas}{Rakotoarivony}

\address{
    Thales, cortAIx Labs, 1 Av. Augustin Fresnel, 91120 Palaiseau, France
}

\email{lucas.rakotoarivony@thalesgroup.com}


\usepackage{comment}

\begin{document}

\maketitle

\begin{abstract}
Quantization has become essential for the efficient deployment of speech processing systems. Although widely studied, most existing quantization methods were developed for vision and NLP architectures, while the specific challenges of audio signals remain largely overlooked. In particular, we show that audio activations can exhibit large calibration ranges, leading to significant information loss when standard calibration techniques are applied.
To address this, we propose ESC, an Evolution Strategy-based Calibration method that formulates activation scaling as an optimization problem and solves it using a two-step local-global scheme driven by an evolution strategy.
ESC enables unaltered performance under full INT8 quantization and is the first calibration method to achieve near-lossless performance for full INT4 quantization across multiple speech tasks. Integrating ESC with PTQ methods further reduces performance loss, achieving a 1\% relative accuracy degradation on the AST model.

\end{abstract}

\section{Introduction}

Modern speech models have achieved near human-level performance on many tasks thanks to large-scale pretraining on massive datasets \cite{radford2023robust,baevski2020wav2vec} and advanced architectures like transformers-based models \cite{gulati2020conformer,gong2021ast}. However, deploying these models in real-world scenarios with limited memory and computational resources typically requires quantization into hardware-friendly integer formats.

Quantization \cite{wang2024model} is widely used for low-bit neural network deployment because it reduces numerical precision of weights and activations, enabling faster inference with integer operations and lowering memory and storage costs. Although quantization has been extensively studied in computer vision \cite{nagel2020up,li2021brecq} and natural language processing (NLP) \cite{xiao2023smoothquant,frantar2022gptq}, the audio domain remains underexplored.
Most existing audio work \cite{li2025towards,kawamura2025bittts} relies on quantization-aware training (QAT) \cite{jacob2018quantization}, which requires access to a non-negligible portion of the training data. Some studies \cite{wagner2024outlier,gu2025ultra,shao2023whisper} investigate post-training quantization (PTQ) for speech models, but they mainly target specific architectures \cite{wagner2024outlier,shao2023whisper} or focus on weight quantization \cite{gu2025ultra}, often neglecting activation quantization, which is essential for fully integer inference. As a result, a complete integer quantization pipeline for general speech models remains an open problem.

\begin{figure}[t]
  \centering
   \includegraphics[width=1.0\linewidth]{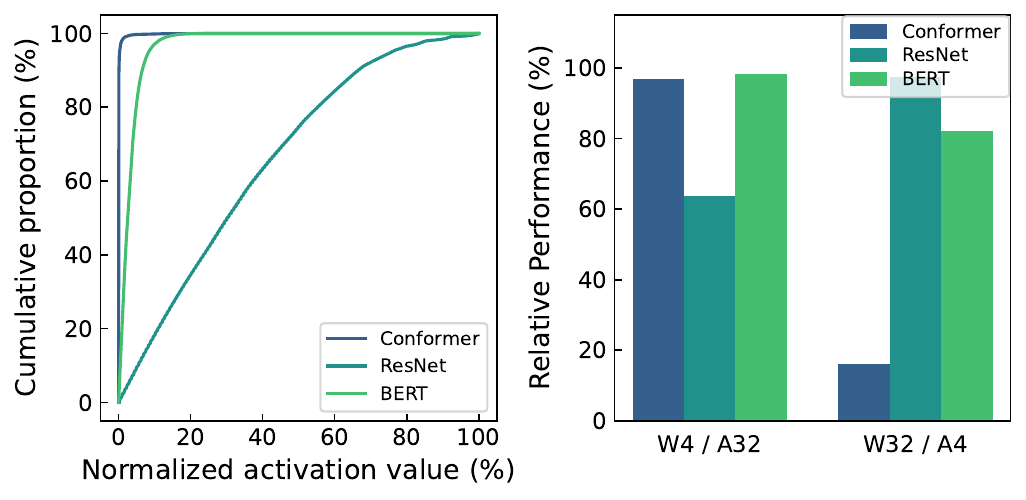}
   \caption{Illustration of quantization behavior across audio (Conformer \cite{gulati2020conformer}), vision (ResNet \cite{he2016deep}), and NLP (BERT \cite{devlin2019bert}) models. \textbf{Left}: Cumulative distribution of normalized activation values, showing an approximately uniform distribution for ResNet, a rapidly saturating distribution for BERT, and a highly compressed distribution for Conformer. \textbf{Right}: Relative performance under weight and activation quantization using max calibration. While all models maintain good performance with 4-bit weight quantization, 4-bit activation quantization severely degrades performance for Conformer, unlike ResNet and BERT.}
   \label{fig:motivation}
\end{figure}

This context highlights the need for quantization techniques tailored to the characteristics of audio signals that preserve model performance while substantially reducing model size. As shown in the left part of \Cref{fig:motivation}, audio activations can exhibit extremely large dynamic ranges, unlike typical activations in vision and NLP tasks. Consequently, standard calibration methods \cite{vanhoucke2011improving,mckinstry2018discovering,nvidia8bit,choukroun2019low} that estimate quantization ranges often produce highly unbalanced quantization bins, causing most values to be mapped to the same integer level, leading to severe information loss, as illustrated in the right part of \Cref{fig:motivation}.

Motivated by this challenge, we propose a new calibration method called Evolution Strategy-Based Calibration (ESC), which uses an evolution strategy \cite{rechenberg1978evolutionsstrategien} to optimize activation scaling \cite{gholami2022survey} through an explicit optimization formulation. As shown in \Cref{fig:motivation}, activation calibration is a key difficulty in speech model quantization. We address this by formulating calibration as a two-step optimization process that integrates local and global objectives.
First, scale factors are initialized using an MSE-based approach \cite{choukroun2019low} that minimizes the reconstruction error between FP32 and quantized layer outputs. 
Then, inspired by global optimization methods such as BRECQ \cite{li2021brecq} and QAT \cite{jacob2018quantization}, we formulate the problem as a joint optimization over all activation scale factors and solve it using an evolution strategy to handle its non-smooth and non-differentiable nature.
Experiments on multiple speech tasks show that ESC consistently outperforms existing calibration methods and achieves lossless performance for full INT8 quantization. For INT4 settings, when combined with state-of-the-art PTQ methods \cite{nagel2020up,li2021brecq}, ESC calibration achieves near-lossless quantization while incurring only a modest performance drop and maintaining high accuracy. 
The main contributions of this paper are summarized as follows:

\begin{itemize}
    \item We formulate calibration as a local-global optimization problem and propose a novel calibration scheme that uses an evolution strategy to minimize quantization error.
    \item We conduct extensive experiments demonstrating the superiority of ESC over standard calibration schemes and showing minimal performance degradation across various models.
    \item We deploy the quantized models and observe an average inference speedup of 2.31$\times$, along with a substantial reduction in memory usage.
\end{itemize}

\section{Related Work}

\subsection{Quantization} \label{subsec:quantization}

Quantization enables deployment of neural networks in resource-constrained settings by reducing memory and computational requirements \cite{wang2024model}. It maps floating-point values to discrete integers while aiming to preserve model accuracy. A key challenge, called calibration, is the selection of the scaling factor, which is determined by a clipping range that truncates real-valued inputs and influences model performance. Traditional min-max statistics \cite{vanhoucke2011improving} are sensitive to outliers, while alternatives include percentile-based calibration \cite{mckinstry2018discovering} or optimization-based criteria \cite{nvidia8bit,choukroun2019low}. PTQ further mitigates accuracy loss without retraining and has been studied for vision models, such as CNNs \cite{nagel2020up,li2021brecq}, ViTs \cite{liu2023noisyquant,kim2024hyq}, and diffusion models \cite{dong2025ditas}, as well as for language models \cite{xiao2023smoothquant,gong2024minimize}. Quantization for speech models, however, remains relatively underexplored \cite{kawamura2025bittts,gu2025ultra,shao2023whisper}.

\subsection{Evolution Strategies} \label{subsec:evolution}

Evolutionary algorithms (EA) \cite{sloss20192019} are stochastic, population-based optimizers inspired by natural evolution. They iteratively improve candidate solutions through selection, evolution, and evaluation, enabling effective approximation of optima for complex problems. Evolution Strategies (ES) \cite{rechenberg1978evolutionsstrategien}, a common EA subclass for continuous optimization, rely on mutation of real-valued parameters and self-adaptive step sizes. Notable ES variants include estimation-of-distribution methods such as CMA-ES \cite{hansen2006cma}, natural evolution strategies \cite{schaul2011high,wierstra2014natural}, and finite-difference methods like OpenAI-ES \cite{salimans2017evolution}.

\section{Methods}

\subsection{Quantization Formulation}

As proposed in \cite{gholami2022survey}, we employ the widely used quantization scheme defined as follows:

\begin{equation}
\label{eq:quantization_formula}
Q(r) = \text{Int}\big({r}/{s}\big)-Z,
\end{equation}

Here, $Q$ denotes the quantization operator, $r$ is a real-valued input (either an activation or a weight), $s$ is a real-valued scaling factor, and $Z$ is an integer zero point. The $\text{Int}$ function maps a real value to an integer through a rounding operation. This approach is referred to as uniform quantization, as the resulting quantization levels are evenly spaced. We adopt this strategy for hardware-friendly compatibility, since non-uniform quantization schemes are typically challenging to implement efficiently on general-purpose hardware, such as GPUs and CPUs \cite{gholami2022survey}.
A critical factor in this process is the selection of the scaling factor $s$. As proposed in \cite{gholami2022survey}, $s$ is defined as:

\begin{equation}
    s = \frac{\beta - \alpha}{2^{b} - 1},
\end{equation}

Here, $[\alpha, \beta]$ represents the clipping range, an interval used to clip the real-valued inputs, and $b$ is the quantization bit width. Consequently, defining the scaling factor requires first determining the clipping range. We specifically consider the case where $\alpha=-\beta$, referred to as symmetric quantization, which simplifies the process by setting the zero point $Z$ to $0$.

The process of selecting the clipping range, and consequently the scaling factor $s$, is commonly referred to as calibration. Weight calibration is relatively straightforward, as the weights of a trained model are fixed and typically follow a Gaussian distribution \cite{yu2019gdrq,strom2022squashed}. In contrast, activation distributions are more sensitive and can vary substantially across samples. Therefore, activation calibration generally requires $n$ samples to estimate a representative activation distribution.
Standard strategies set $\beta$ either to the maximum observed value or to a chosen percentile of the activation distribution.

\begin{figure}[t]
  \centering
   \includegraphics[width=1.0\linewidth]{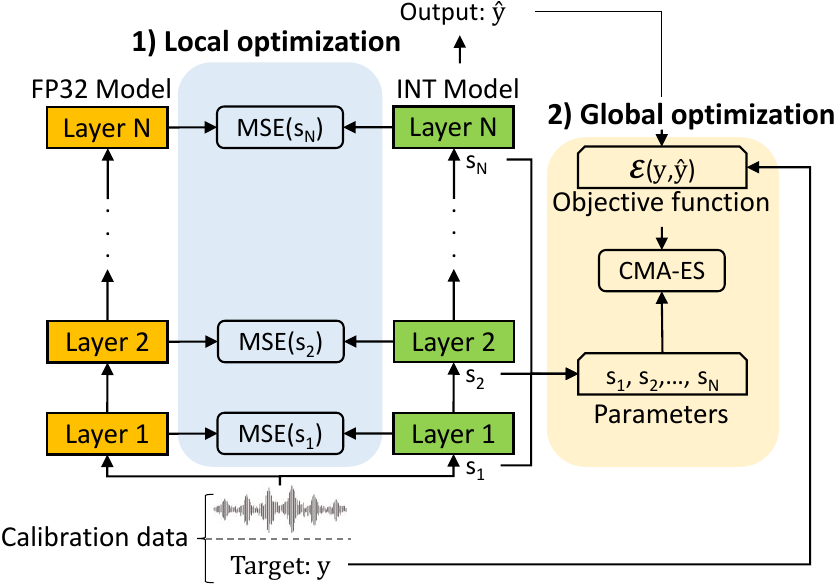}
   \caption{\textbf{Overview of the proposed ESC method}. First, each layer-wise activation scaling factor is locally optimized by minimizing the MSE between the FP32 and quantized layer outputs. Then, all scaling factors are jointly refined using the CMA-ES algorithm to minimizes the task-specific error between the quantized model output $\hat{y}$ and the target $y$.}
   \label{fig:overview}
\end{figure}

\begin{table*}[t]
\caption{Comparison of calibration methods across multiple models and speech tasks for fully 4-bit and 8-bit quantized models. The symbols $\downarrow$ and $\uparrow$ indicate that lower and higher values are better, respectively. Best results are shown in bold.}
	\centering
	\begin{tabular}{@{}cccccccccccc@{}}
		\toprule
		\multirow{2}{*}{Method} & \multirow{2}{*}{Bits} & \multicolumn{2}{c}{Conformer} &  \multicolumn{2}{c}{ECAPA} & \multicolumn{2}{c}{MP-SENet} & \multicolumn{2}{c}{FastSpeech 2} & \multicolumn{2}{c}{AST} \\ \cmidrule(l){3-4} \cmidrule(l){5-6} \cmidrule(l){7-8} \cmidrule(l){9-10} \cmidrule(l){11-12}
        & & WER $\downarrow$ & CER $\downarrow$ & EER $\downarrow$ & minDCF $\downarrow$ & PESQ $\uparrow$ & STOI $\uparrow$ & Mel $\downarrow$ & PostNet $\downarrow$ & Acc $\uparrow$ & mAP $\uparrow$ \\ \midrule
        Full precision & 32 & 15.94 & 4.84 & 0.97 & 7.17 & 2.12 & 93.67 & 40.18 & 40.09 & 98.13 & 99.98 \\ \midrule
        Max \cite{vanhoucke2011improving} & 8 & 16.54 & 5.04 & 1.11 & 7.99 & \textbf{2.16} & 93.31 & 40.88 & 40.72 & 98.12 & \textbf{99.98} \\
        Percentile \cite{mckinstry2018discovering} & 8 & 16.07 & 4.85 & 0.99 & \textbf{7.06} & 2.11 & 92.88 & 40.55 & 40.58 & 98.06 & 99.95 \\
        Entropy \cite{nvidia8bit} & 8 & 16.27 & 4.93 & 18.70 & 78.33 & 2.11 & \textbf{93.55} & 161.48 & 161.59 & 90.09 & 99.65 \\
        MSE  \cite{choukroun2019low} & 8 & 16.09 & 4.87 & 0.99 & 7.53 & 2.09 & 92.72 & 40.45 & 40.49 & 98.07 & 99.96 \\
        ESC & 8 & \textbf{16.01} & \textbf{4.83} & \textbf{0.94} & 7.65 & 2.11 & 92.90 & \textbf{40.35} & \textbf{40.32} & \textbf{98.15} & 99.96 \\ \midrule
        Max \cite{vanhoucke2011improving} & 4 & 144.14 & 84.81 & 44.40 & 99.99 & 1.16 & 67.52 & 231.40 & 231.25 & 3.71 & 50.00 \\
        Percentile \cite{mckinstry2018discovering} & 4 & 50.83 & 18.60 & 26.42 & 97.49 & 2.22 & 90.10 & 138.99 & 138.98 & 95.51 & 99.90 \\
        Entropy \cite{nvidia8bit} & 4 & 50.83 & 18.76 & 19.44 & 86.12 & 2.44 & 93.64 & 297.44 & 297.60 & 63.98 & 96.15 \\
        MSE \cite{choukroun2019low} & 4 & 41.22 & 14.61 & 13.07 & 65.66 & 2.50 & 92.99 & 130.56 & 130.41 & 96.03 & 99.91 \\
        ESC & 4 & \textbf{38.49} & \textbf{13.50} & \textbf{11.28} & \textbf{63.05} & \textbf{2.51} & \textbf{93.79} & \textbf{98.34} & \textbf{98.27} & \textbf{96.41} & \textbf{99.94} \\
		\bottomrule[0.5pt]     
	\end{tabular}
	\label{tab:calibration}
\end{table*}

\subsection{Scale Initialization}

Since audio activations can exhibit extremely large dynamic ranges, standard calibration strategies such as Max \cite{vanhoucke2011improving} or Percentile \cite{mckinstry2018discovering} often produce poorly distributed quantization bins, leading to significant information loss, as illustrated in \Cref{fig:motivation}. 
To address this, rather than selecting the scaling factor solely based on the activation distribution, we cast the calibration process as an optimization problem that directly minimizes the task-specific error, as defined below:

\begin{equation}
\label{eq:objective}
\mathbf{S}^\ast = 
\arg\min_{\mathbf{S}}
\mathcal{E}(f_q(x;\mathbf{S}),\, y),
\end{equation}

where $f_q(\cdot)$ denotes the already trained quantized neural network with fixed weights, $\mathbf{S} = \{s_1, \dots, s_N\}$ is the vector of per-layer activation scale parameters, $x$ represents the calibration samples, $y$ are the reference targets, and $\mathcal{E}(\cdot)$ is the task-dependent error metric.  
To initialize $\mathbf{S}$, we adopt the MSE-based calibration algorithm \cite{choukroun2019low}, which treats the optimization as a local procedure. Specifically, for each layer $i = 1, \dots, N$, the algorithm optimizes its activation scale $s_i$ independently:  

\begin{equation}
s_i^\ast = \arg\min_{s_i} \lVert l(x) - l_q(x; s_i) \rVert_2^2,
\end{equation}

where $l$ is the layer of interest and $l_q$ is its quantized counterpart. Collectively, the resulting scales $\mathbf{S}$ provide a stable initialization point for subsequent global refinement.

\subsection{Evolution Strategy Optimization}

Inspired by BRECQ \cite{li2021brecq} and QAT \cite{jacob2018quantization} approaches, we argue that optimizing quantization scales only locally is insufficient to achieve optimal performance, as such methods do not account for cross-layer dependencies. To address this limitation, we propose to optimize the scale vector $\mathbf{S}$ globally using an evolution strategy, with the objective of minimizing the loss defined in \Cref{eq:objective}.
As discussed in \Cref{subsec:evolution}, evolution strategies are well suited for the optimization of multiple continuous variables in settings where the objective function is non-convex and non-differentiable. In particular, we adopt the CMA-ES \cite{hansen2006cma} algorithm, which provides a robust gradient-free optimization framework for continuous parameters and has demonstrated strong performance on ill-conditioned and noisy objective functions \cite{hansen2006cma,loshchilov2016cma}.

At each iteration $t$, CMA-ES samples a population of $\lambda$ candidate scale vectors from a multivariate normal distribution:

\begin{equation}
\mathbf{S}^{(t)}_k \sim \mathcal{N}\bigl(\mathbf{m}^{(t)},\, \sigma_t^2 \mathbf{C}^{(t)}\bigr),
\quad k = 1, \dots, \lambda.
\end{equation}

where $\mathbf{m}^{(t)}$ denotes the mean of the search distribution, $\mathbf{C}^{(t)}$ is the covariance matrix encoding parameter correlations, and $\sigma_t$ is the global step size controlling the exploration radius. Each sampled candidate $\mathbf{S}^{(t)}_k$ is evaluated using the objective defined in \Cref{eq:objective}. Based on the ranking of these candidates, the parameters $(\mathbf{m}^{(t)}, \mathbf{C}^{(t)}, \sigma_t)$ are updated following the standard CMA-ES update rules.
The optimization process terminates when the total number of objective function evaluations reaches a predefined budget $\Gamma$.

To obtain the final optimized scaling vector $\mathbf{S}$, we use the mean of the final sampling distribution produced by the CMA-ES algorithm instead of the single best-evaluated solution, in order to improve robustness \cite{hansen2006cma}.
An overview of the proposed ESC method is provided in \Cref{fig:overview}.


\section{Experiments}

\begin{table*}[t]
\caption{Evaluation of PTQ techniques from the NLP and vision domains applied on top of ESC calibration.}
	\centering
	\begin{tabular}{@{}cccccccccccc@{}}
		\toprule
		\multirow{2}{*}{Method} & \multirow{2}{*}{Bits} & \multicolumn{2}{c}{Conformer} &  \multicolumn{2}{c}{ECAPA} & \multicolumn{2}{c}{MP-SENet} & \multicolumn{2}{c}{FastSpeech 2} & \multicolumn{2}{c}{AST} \\ \cmidrule(l){3-4} \cmidrule(l){5-6} \cmidrule(l){7-8} \cmidrule(l){9-10} \cmidrule(l){11-12}
        & & WER $\downarrow$ & CER $\downarrow$ & EER $\downarrow$ & minDCF $\downarrow$ & PESQ $\uparrow$ & STOI $\uparrow$ & Mel $\downarrow$ & PostNet $\downarrow$ & Acc $\uparrow$ & mAP $\uparrow$ \\ \midrule
        Full precision & 32 & 15.94 & 4.84 & 0.97 & 7.17 & 2.12 & 93.67 & 40.18 & 40.09 & 98.13 & 99.98 \\ \midrule
        ESC & 4 & \textbf{38.49} & \textbf{13.50} & 11.28 & 63.05 & 2.51 & 93.79 & 98.34 & 98.27 & 96.41 & 99.94 \\ \midrule
        Adaround \cite{nagel2020up} & 4 & 106.79 & 75.44 & 21.76 & 89.39 & 2.45 & 92.67 & \textbf{83.53} & \textbf{83.46} & 88.53 & 99.70 \\
        NoisyQuant \cite{liu2023noisyquant} & 4 & 39.89 & 13.96 & 11.28 & 63.05 & 2.50 & 93.61 & 95.80 & 95.73 & 96.26 & 99.94 \\
        DiTAS \cite{dong2025ditas} & 4 & 99.24 & 84.58 & 11.68 & 66.00 & 2.51 & 94.51 & 98.34 & 98.27 & 96.40 & 99.94 \\
        HyQ \cite{kim2024hyq} & 4 & 98.01 & 70.43 & \textbf{8.20} & \textbf{49.70} & 2.16 & 93.31 & 99.19 & 99.02 & \textbf{96.76} & \textbf{99.95} \\
        BRECQ \cite{li2021brecq} & 4 & 64.17 & 27.66 & 12.01 & 62.45 & 2.21 & \textbf{96.83} & 84.59 & 84.45 & 96.17 & 99.89 \\
        SmoothQuant \cite{xiao2023smoothquant} & 4 & 39.02 & 13.60 & 11.28 & 63.05 & \textbf{2.54} & 96.32 & 98.34 & 98.27 & 94.38 & 99.88 \\
        BC \cite{gong2024minimize} & 4 & 38.57 & 13.58 & 11.27 & 63.87 & 2.53 & 93.90 & 90.21 & 90.14 & 96.23 & 99.91 \\
		\bottomrule[0.5pt]     
	\end{tabular}
	\label{tab:ptq}
\end{table*}

\subsection{Experimental Setup}

To ensure a representative evaluation across speech processing domains, we conduct experiments on five widely used speech-based tasks: speech recognition, speaker recognition, speech enhancement, text-to-speech, and audio classification. For each task, the corresponding model is evaluated on the official test split of the associated dataset.

\begin{itemize}
    \item \textbf{Speech recognition}: Conformer \cite{gulati2020conformer} on LibriSpeech \cite{panayotov2015librispeech}, evaluated using Word Error Rate (WER) and Character Error Rate (CER).
    \item \textbf{Speaker recognition}: ECAPA \cite{desplanques2020ecapa} on VoxCeleb \cite{chung2018voxceleb2}, evaluated using Equal Error Rate (EER) and minimum Detection Cost Function (minDCF).
    \item \textbf{Speech enhancement}: MP-SENet \cite{lu2023mp} on VoiceBank-DEMAND \cite{botinhao2016investigating}, evaluated using Perceptual Evaluation of Speech Quality (PESQ) and Short-Time Objective Intelligibility (STOI).
    \item \textbf{Text-to-speech}: FastSpeech 2 \cite{ren2020fastspeech} on LJSpeech \cite{ljspeech17}, evaluated using Mel-spectrogram loss and PostNet loss.
    \item \textbf{Audio classification}: AST \cite{gong2021ast} on Speech Commands V2 \cite{warden2018speech}, evaluated using Accuracy (Acc) and mean Average Precision (mAP).
\end{itemize}

All evaluated models are fully quantized to either INT8 or INT4 precision, with both weights and activations quantized to enable a completely integer-only inference pipeline. For both calibration and the evolution strategy, we use the same $n=100$ samples from the training set for each task.
To determine the scaling factors for activation quantization of the Convolutional, Linear, and LayerNorm operators, we apply our proposed method ESC. For all other operators, as well as for weight quantization, we employ the Max calibration strategy.
For the CMA-ES algorithm, the initial step size is set to $\sigma=0.1$ and the predefined budget to
$\Gamma=100$. All experiments, including calibration, evolution strategy, PTQ, and inference, are conducted on a NVIDIA RTX 3090 GPU.

\subsection{Comparison with Baseline Calibration Methods} 

We evaluate several popular calibration methods across multiple tasks and models using the pytorch\_quantization \cite{pytorch_quantization} framework, which provides implementations of Max \cite{vanhoucke2011improving}, Percentile \cite{mckinstry2018discovering}, Entropy \cite{nvidia8bit}, and MSE \cite{choukroun2019low} calibration. For the Percentile method, we test the 99.99, 99.999, and 99.9999 percentiles and report the best result for simplicity.

As shown in \Cref{tab:calibration}, the proposed ESC method consistently outperforms baseline strategies, especially in INT4 quantization. Max performs reasonably in INT8 but drops in INT4 due to outliers compressing the activation distribution. Percentile often achieves strong results by removing extreme outliers, but its optimal percentile varies across models and tasks, requiring extra tuning. Entropy works well for some architectures like Conformer and ECAPA, but can significantly degrade performance for others, such as FastSpeech 2 and AST, even in INT8. Among baselines, MSE provides the best overall performance, and since our initialization uses MSE-derived scaling factors, applying ESC further improves results. On average, ESC achieves the best INT8 performance and substantial gains across all INT4 scenarios.

Interestingly, quantization can even improve performance for some models. For MP-SENet, ESC in INT4 achieves a PESQ of 2.51, a 18\% relative improvement over FP32, likely due to the regularizing effect of quantization, which suppresses minor weight contributions and stabilizes outputs for cleaner speech. For AST, ESC in INT4 causes only a 1.75\% relative accuracy drop, showing minimal degradation under fully INT4 quantization.

\subsection{Integration with State-of-the-Art PTQ Methods}

Since our method is a calibration strategy for determining activation scaling factors, it can be easily combined with existing PTQ techniques. As discussed in \Cref{subsec:quantization}, PTQ for speech models remains relatively underexplored. However, many PTQ methods developed for vision and NLP \cite{nagel2020up,li2021brecq,xiao2023smoothquant,gong2024minimize,liu2023noisyquant,kim2024hyq,dong2025ditas} can be directly applied to speech models.

\Cref{tab:ptq} presents the performance of various PTQ strategies in the audio domain when combined with ESC calibration. Overall, no single method consistently outperforms the others, as results vary significantly across different models. Some approaches that perform well on certain models can substantially degrade performance on others, highlighting the need for PTQ methods specifically tailored to the audio domain.

Methods such as NoisyQuant \cite{liu2023noisyquant}, BC \cite{gong2024minimize}, and SmoothQuant \cite{xiao2023smoothquant} generally provide small but consistent improvements over the baseline without significant degradation, showing that techniques transferred from other domains can benefit speech models. Larger gains are observed in specific cases: HyQ \cite{kim2024hyq} applied to ECAPA and Adaround \cite{nagel2020up} applied to FastSpeech 2 improve performance by 27\% and 15\%, respectively, relative to the ESC baseline. In addition, combining AST with HyQ achieves 96.76\% accuracy, approaching FP32 performance. These results indicate that, when paired with suitable PTQ strategies, ESC calibration can enable near-lossless quantization.

\begin{table}[t]
\caption{Comparison of inference latency and model size between FP32 and INT8 versions of several speech models.}
\begin{tabular}{@{}c|ccc|cc@{}}
\toprule
\multirow{2}{*}{Model} & \multicolumn{2}{c}{Latency (ms)} & \multirow{2}{*}{Speedup} & \multicolumn{2}{c}{Size (MB)} \\ \cmidrule(l){2-3} \cmidrule(l){5-6}
                       & FP32         & INT8         &    & FP32 & INT8                     \\ \midrule
Conformer              &  7.27            &   5.42          & 1.34$\times$                 & 112.63 &  43.64      \\
ECAPA                  &  2.19            &  1.03            &  2.13$\times$                   & 63.58 & 59.69     \\
MP-SENet               &  55.95            &  38.35            &   1.46$\times$ & 23.86 & 9.15                     \\
FastSpeech 2            &  23.86            & 15.45             & 1.54$\times$                       & 398.70 & 216.12  \\
AST                    &  25.11            & 4.95             &   5.07$\times$                    & 331.69 & 113.45 \\
\bottomrule[0.5pt]
\end{tabular}
\label{tab:latency}
\end{table}

\subsection{Latency and Model Size Evaluation}

We evaluate inference latency and memory footprint of quantized speech models deployed on an NVIDIA GeForce RTX 3090 GPU, which provides Tensor Cores supporting accelerated INT8 execution. Models are exported using TorchScript \cite{paszke2019pytorch} tracing and deployed with TensorRT \cite{tensorrt}. Although this GPU is chosen for its mature software ecosystem \cite{paszke2019pytorch,tensorrt} and reliable deployment tools, our quantization and export strategy is not limited to GPUs and can be extended to other hardware platforms, including embedded AI processors \cite{hager202411,reuther2020survey}. 
Since Tensor Cores accelerate INT8 operations for both weights and activations, but only accelerate INT4 operations for weights, we restrict our experiments to 8-bit quantized models. 
\Cref{tab:latency} compares FP32 and INT8 implementations in terms of latency and model size. Results show that INT8 models consistently reduce memory usage and achieve notable speedups, ranging from 1.34$\times$ to 5.07$\times$.

\section{Conclusion}

In this paper, we proposed a novel two-stage calibration scheme that combines local MSE-based optimization with a global evolutionary strategy to optimize activation scaling factors in speech models. Our study highlights that, unlike in vision or NLP, audio models are particularly sensitive to activation quantization, providing strong motivation for our approach. Experimental results demonstrate that our method preserves full-precision performance for fully 8-bit models while achieving an average inference speedup of 2.31$\times$. Moreover, when applied to fully 4-bit models in combination with modern PTQ techniques, our calibration scheme delivers near-lossless performance across a wide range of speech models and tasks.

\newpage

\bibliographystyle{IEEEtran}
\bibliography{esc}

@article{wang2024model,
  title={Model compression and efficient inference for large language models: A survey},
  author={Wang, Wenxiao and Chen, Wei and Luo, Yicong and Long, Yongliu and Lin, Zhengkai and Zhang, Liye and Lin, Binbin and Cai, Deng and He, Xiaofei},
  journal={arXiv preprint arXiv:2402.09748},
  year={2024}
}

@article{mckinstry2018discovering,
  title={Discovering low-precision networks close to full-precision networks for efficient embedded inference},
  author={McKinstry, Jeffrey L and Esser, Steven K and Appuswamy, Rathinakumar and Bablani, Deepika and Arthur, John V and Yildiz, Izzet B and Modha, Dharmendra S},
  journal={arXiv preprint arXiv:1809.04191},
  year={2018}
}

@article{nvidia8bit,
  title={Nvidia 8-bit inference with TensorRT},
  author={Migacz, Szymon},
  journal={GPU Technology Conference},
  year={2017}
}

@inproceedings{vanhoucke2011improving,
  title={Improving the speed of neural networks on CPUs},
  author={Vanhoucke, Vincent and Senior, Andrew and Mao, Mark Z and others},
  booktitle={Proc. deep learning and unsupervised feature learning NIPS workshop},
  volume={1},
  number={2011},
  pages={4},
  year={2011}
}

@inproceedings{choukroun2019low,
  title={Low-bit quantization of neural networks for efficient inference},
  author={Choukroun, Yoni and Kravchik, Eli and Yang, Fan and Kisilev, Pavel},
  booktitle={2019 IEEE/CVF International Conference on Computer Vision Workshop (ICCVW)},
  pages={3009--3018},
  year={2019},
  organization={IEEE}
}

@inproceedings{nagel2020up,
  title={Up or down? adaptive rounding for post-training quantization},
  author={Nagel, Markus and Amjad, Rana Ali and Van Baalen, Mart and Louizos, Christos and Blankevoort, Tijmen},
  booktitle={International conference on machine learning},
  pages={7197--7206},
  year={2020},
  organization={PMLR}
}

@inproceedings{liu2023noisyquant,
  title={Noisyquant: Noisy bias-enhanced post-training activation quantization for vision transformers},
  author={Liu, Yijiang and Yang, Huanrui and Dong, Zhen and Keutzer, Kurt and Du, Li and Zhang, Shanghang},
  booktitle={Proceedings of the IEEE/CVF Conference on Computer Vision and Pattern Recognition},
  pages={20321--20330},
  year={2023}
}

@inproceedings{dong2025ditas,
  title={Ditas: Quantizing diffusion transformers via enhanced activation smoothing},
  author={Dong, Zhenyuan and Zhang, Sai Qian},
  booktitle={2025 IEEE/CVF Winter Conference on Applications of Computer Vision (WACV)},
  pages={4606--4615},
  year={2025},
  organization={IEEE}
}

@inproceedings{kim2024hyq,
  title={Hyq: Hardware-friendly post-training quantization for cnn-transformer hybrid networks},
  author={Kim, Nam Joon and Lee, Jongho and Kim, Hyun},
  booktitle={Proceedings of the Thirty-Third International Joint Conference on Artificial Intelligence, IJCAI-24. International Joint Conferences on Artificial Intelligence Organization},
  volume={8},
  pages={4291--4299},
  year={2024}
}

@article{li2021brecq,
  title={Brecq: Pushing the limit of post-training quantization by block reconstruction},
  author={Li, Yuhang and Gong, Ruihao and Tan, Xu and Yang, Yang and Hu, Peng and Zhang, Qi and Yu, Fengwei and Wang, Wei and Gu, Shi},
  journal={arXiv preprint arXiv:2102.05426},
  year={2021}
}

@inproceedings{xiao2023smoothquant,
  title={Smoothquant: Accurate and efficient post-training quantization for large language models},
  author={Xiao, Guangxuan and Lin, Ji and Seznec, Mickael and Wu, Hao and Demouth, Julien and Han, Song},
  booktitle={International conference on machine learning},
  pages={38087--38099},
  year={2023},
  organization={PMLR}
}

@article{gong2024minimize,
  title={Minimize Quantization Output Error with Bias Compensation},
  author={Gong, Cheng and Zheng, Haoshuai and Hu, Mengting and Lin, Zheng and Fan, Deng-Ping and Zhang, Yuzhi and Li, Tao},
  journal={arXiv preprint arXiv:2404.01892},
  year={2024}
}

@inproceedings{rechenberg1978evolutionsstrategien,
  title={Evolutionsstrategien},
  author={Rechenberg, Ingo},
  booktitle={Simulationsmethoden in der Medizin und Biologie: Workshop, Hannover, 29. Sept.--1. Okt. 1977},
  pages={83--114},
  year={1978},
  organization={Springer}
}

@article{sloss20192019,
  title={2019 evolutionary algorithms review},
  author={Sloss, Andrew N and Gustafson, Steven},
  journal={arXiv preprint arXiv:1906.08870},
  year={2019}
}

@article{hansen2006cma,
  title={The CMA evolution strategy: a comparing review},
  author={Hansen, Nikolaus},
  journal={Towards a new evolutionary computation: Advances in the estimation of distribution algorithms},
  pages={75--102},
  year={2006},
  publisher={Springer}
}

@inproceedings{schaul2011high,
  title={High dimensions and heavy tails for natural evolution strategies},
  author={Schaul, Tom and Glasmachers, Tobias and Schmidhuber, J{\"u}rgen},
  booktitle={Proceedings of the 13th annual conference on Genetic and evolutionary computation},
  pages={845--852},
  year={2011}
}

@article{wierstra2014natural,
  title={Natural evolution strategies},
  author={Wierstra, Daan and Schaul, Tom and Glasmachers, Tobias and Sun, Yi and Peters, Jan and Schmidhuber, J{\"u}rgen},
  journal={The Journal of Machine Learning Research},
  volume={15},
  number={1},
  pages={949--980},
  year={2014},
  publisher={JMLR. org}
}

@article{salimans2017evolution,
  title={Evolution strategies as a scalable alternative to reinforcement learning},
  author={Salimans, Tim and Ho, Jonathan and Chen, Xi and Sidor, Szymon and Sutskever, Ilya},
  journal={arXiv preprint arXiv:1703.03864},
  year={2017}
}

@article{gulati2020conformer,
  title={Conformer: Convolution-augmented transformer for speech recognition},
  author={Gulati, Anmol and Qin, James and Chiu, Chung-Cheng and Parmar, Niki and Zhang, Yu and Yu, Jiahui and Han, Wei and Wang, Shibo and Zhang, Zhengdong and Wu, Yonghui and others},
  journal={arXiv preprint arXiv:2005.08100},
  year={2020}
}

@article{lu2023mp,
  title={MP-SENet: A speech enhancement model with parallel denoising of magnitude and phase spectra},
  author={Lu, Ye-Xin and Ai, Yang and Ling, Zhen-Hua},
  journal={arXiv preprint arXiv:2305.13686},
  year={2023}
}

@article{ren2020fastspeech,
  title={Fastspeech 2: Fast and high-quality end-to-end text to speech},
  author={Ren, Yi and Hu, Chenxu and Tan, Xu and Qin, Tao and Zhao, Sheng and Zhao, Zhou and Liu, Tie-Yan},
  journal={arXiv preprint arXiv:2006.04558},
  year={2020}
}

@article{gong2021ast,
  title={Ast: Audio spectrogram transformer},
  author={Gong, Yuan and Chung, Yu-An and Glass, James},
  journal={arXiv preprint arXiv:2104.01778},
  year={2021}
}

@article{desplanques2020ecapa,
  title={Ecapa-tdnn: Emphasized channel attention, propagation and aggregation in tdnn based speaker verification},
  author={Desplanques, Brecht and Thienpondt, Jenthe and Demuynck, Kris},
  journal={arXiv preprint arXiv:2005.07143},
  year={2020}
}

@article{frantar2022gptq,
  title={Gptq: Accurate post-training quantization for generative pre-trained transformers},
  author={Frantar, Elias and Ashkboos, Saleh and Hoefler, Torsten and Alistarh, Dan},
  journal={arXiv preprint arXiv:2210.17323},
  year={2022}
}

@article{li2025towards,
  title={Towards One-bit ASR: Extremely Low-bit Conformer Quantization Using Co-training and Stochastic Precision},
  author={Li, Zhaoqing and Xu, Haoning and Jin, Zengrui and Meng, Lingwei and Wang, Tianzi and Wang, Huimeng and Chen, Youjun and Cui, Mingyu and Hu, Shujie and Liu, Xunying},
  journal={arXiv preprint arXiv:2505.21245},
  year={2025}
}

@article{kawamura2025bittts,
  title={BitTTS: Highly Compact Text-to-Speech Using 1.58-bit Quantization and Weight Indexing},
  author={Kawamura, Masaya and Hasumi, Takuya and Shirahata, Yuma and Yamamoto, Ryuichi},
  journal={arXiv preprint arXiv:2506.03515},
  year={2025}
}

@inproceedings{jacob2018quantization,
  title={Quantization and training of neural networks for efficient integer-arithmetic-only inference},
  author={Jacob, Benoit and Kligys, Skirmantas and Chen, Bo and Zhu, Menglong and Tang, Matthew and Howard, Andrew and Adam, Hartwig and Kalenichenko, Dmitry},
  booktitle={Proceedings of the IEEE conference on computer vision and pattern recognition},
  pages={2704--2713},
  year={2018}
}

@article{wagner2024outlier,
  title={Outlier Reduction with Gated Attention for Improved Post-training Quantization in Large Sequence-to-sequence Speech Foundation Models},
  author={Wagner, Dominik and Baumann, Ilja and Riedhammer, Korbinian and Bocklet, Tobias},
  journal={arXiv preprint arXiv:2406.11022},
  year={2024}
}

@inproceedings{gu2025ultra,
  title={Ultra-Low Bit Post-Training Quantization of Large Speech Models via K-Means Clustering and Mixed Precision Allocation},
  author={Gu, Tianteng and Liu, Bei and Wang, Haoyu and Qian, Yanmin},
  booktitle={Proc. Interspeech 2025},
  pages={1988--1992},
  year={2025}
}

@article{shao2023whisper,
  title={Whisper-kdq: A lightweight whisper via guided knowledge distillation and quantization for efficient asr},
  author={Shao, Hang and Wang, Wei and Liu, Bei and Gong, Xun and Wang, Haoyu and Qian, Yanmin},
  journal={CoRR},
  year={2023}
}

@incollection{gholami2022survey,
  title={A survey of quantization methods for efficient neural network inference},
  author={Gholami, Amir and Kim, Sehoon and Dong, Zhen and Yao, Zhewei and Mahoney, Michael W and Keutzer, Kurt},
  booktitle={Low-power computer vision},
  pages={291--326},
  year={2022},
  publisher={Chapman and Hall/CRC}
}

@misc{ljspeech17,
  author       = {Keith Ito and Linda Johnson},
  title        = {The LJ Speech Dataset},
  howpublished = {\url{https://keithito.com/LJ-Speech-Dataset/}},
  year         = 2017
}

@article{warden2018speech,
  title={Speech commands: A dataset for limited-vocabulary speech recognition},
  author={Warden, Pete},
  journal={arXiv preprint arXiv:1804.03209},
  year={2018}
}

@article{chung2018voxceleb2,
  title={Voxceleb2: Deep speaker recognition},
  author={Chung, Joon Son and Nagrani, Arsha and Zisserman, Andrew},
  journal={arXiv preprint arXiv:1806.05622},
  year={2018}
}

@inproceedings{panayotov2015librispeech,
  title={Librispeech: an asr corpus based on public domain audio books},
  author={Panayotov, Vassil and Chen, Guoguo and Povey, Daniel and Khudanpur, Sanjeev},
  booktitle={2015 IEEE international conference on acoustics, speech and signal processing (ICASSP)},
  pages={5206--5210},
  year={2015},
  organization={IEEE}
}

@inproceedings{botinhao2016investigating,
  title={Investigating RNN-based speech enhancement methods for noise-robust text-to-speech},
  author={Botinhao, Cassia Valentini and Wang, Xin and Takaki, Shinji and Yamagishi, Junichi},
  booktitle={9th ISCA speech synthesis workshop},
  pages={159--165},
  year={2016}
}

@inproceedings{he2016deep,
  title={Deep residual learning for image recognition},
  author={He, Kaiming and Zhang, Xiangyu and Ren, Shaoqing and Sun, Jian},
  booktitle={Proceedings of the IEEE conference on computer vision and pattern recognition},
  pages={770--778},
  year={2016}
}

@inproceedings{devlin2019bert,
  title={Bert: Pre-training of deep bidirectional transformers for language understanding},
  author={Devlin, Jacob and Chang, Ming-Wei and Lee, Kenton and Toutanova, Kristina},
  booktitle={Proceedings of the 2019 conference of the North American chapter of the association for computational linguistics: human language technologies, volume 1 (long and short papers)},
  pages={4171--4186},
  year={2019}
}

@ONLINE{tensorrt,
  author={{NVIDIA}},
  title = {Tensor{RT}: https://developer.nvidia.com/tensorrt}
}

@article{paszke2019pytorch,
  title={Pytorch: An imperative style, high-performance deep learning library},
  author={Paszke, Adam and Gross, Sam and Massa, Francisco and Lerer, Adam and Bradbury, James and Chanan, Gregory and Killeen, Trevor and Lin, Zeming and Gimelshein, Natalia and Antiga, Luca and others},
  journal={Advances in neural information processing systems},
  volume={32},
  year={2019}
}

@inproceedings{hager202411,
  title={11.3 Metis AIPU: A 12nm 15TOPS/W 209.6 TOPS SoC for cost-and energy-efficient inference at the edge},
  author={Hager, Pascal Alexander and Moons, Bert and Cosemans, Stefan and Papistas, Ioannis A and Rooseleer, Bram and Van Loon, Jeroen and Uytterhoeven, Roel and Zaruba, Florian and Koumousi, Spyridoula and Stanisavljevic, Milos and others},
  booktitle={2024 IEEE International Solid-State Circuits Conference (ISSCC)},
  volume={67},
  pages={212--214},
  year={2024},
  organization={IEEE}
}

@inproceedings{reuther2020survey,
  title={Survey of machine learning accelerators},
  author={Reuther, Albert and Michaleas, Peter and Jones, Michael and Gadepally, Vijay and Samsi, Siddharth and Kepner, Jeremy},
  booktitle={2020 IEEE high performance extreme computing conference (HPEC)},
  pages={1--12},
  year={2020},
  organization={IEEE}
}

@inproceedings{radford2023robust,
  title={Robust speech recognition via large-scale weak supervision},
  author={Radford, Alec and Kim, Jong Wook and Xu, Tao and Brockman, Greg and McLeavey, Christine and Sutskever, Ilya},
  booktitle={International conference on machine learning},
  pages={28492--28518},
  year={2023},
  organization={PMLR}
}

@article{baevski2020wav2vec,
  title={wav2vec 2.0: A framework for self-supervised learning of speech representations},
  author={Baevski, Alexei and Zhou, Yuhao and Mohamed, Abdelrahman and Auli, Michael},
  journal={Advances in neural information processing systems},
  volume={33},
  pages={12449--12460},
  year={2020}
}

@article{loshchilov2016cma,
  title={CMA-ES for hyperparameter optimization of deep neural networks},
  author={Loshchilov, Ilya and Hutter, Frank},
  journal={arXiv preprint arXiv:1604.07269},
  year={2016}
}

@ONLINE{pytorch_quantization,
  author={{NVIDIA}},
  title = {pytorch\_quantization: http://github.com/NVIDIA/TensorRT}
}

@article{yu2019gdrq,
  title={Gdrq: Group-based distribution reshaping for quantization},
  author={Yu, Haibao and Wen, Tuopu and Cheng, Guangliang and Sun, Jiankai and Han, Qi and Shi, Jianping},
  journal={arXiv preprint arXiv:1908.01477},
  year={2019}
}

@article{strom2022squashed,
  title={Squashed weight distribution for low bit quantization of deep models},
  author={Str{\"o}m, Nikko and Khan, Haidar and Hamza, Wael},
  year={2022}
}

\end{document}